\documentclass[prl,aps,twocolumn,letter] {revtex4}
\begin{document}
In a recent letter~\cite{SZVD}, we demonstrated that there exists a nontrivial correction, arising from the viscosity of the electron liquid, to
the conductance of nanoscale junctions calculated within the
adiabatic local-density approximation (ALDA) to time-dependent
density-functional theory (DFT). This dynamical correction cannot be
captured by any static DFT functional, even the exact one. We showed
that the effect of the viscosity on the conductance can be
analytically predicted, in a qualitative way, using time-dependent
current-density functional theory in the zero-frequency and
linear-response regime~\cite{VUC}. Indeed, in the DC limit, we found
that these viscous effects increase the resistance. In order to provide
an estimate of these effects, we derived Eq. (14) for the dynamical
viscous resistance, and evaluated that expression by employing the
viscosity of bulk electrodes.

Jung {\it et al.} have used our Eq. (14) to calculate the dynamical
resistance of two infinite jellium electrodes separated by a vacuum
gap~\cite{JBG}. By using the local-density dependent formula for the viscosity
coefficient as reported in Ref.~\cite{CV} they conclude that the
dynamical resistance is negligibly small in the specific cases they
consider. Here we argue that their calculations do not preclude the
possibility that the viscosity contribution to the conductance be large in realistic nanoscale structures, the systems of interest
in Ref.~\cite{SZVD}. Two main points support this statement.

1) Our Eq. (14) is an approximate formula, derived under certain
physical assumptions (see below) in order to provide a qualitative
understanding of the viscous effects in nanostructures. As noted in
Ref.~\cite{SZVD}, Eq. (14) was derived assuming homogeneous density
in the transverse direction {\em and} homogeneous current density in
both transverse and longitudinal directions. The only contribution
to the correction we included in Eq. (14) comes from density
variations along the longitudinal direction. It is on the ground of
these assumptions that we adopted the viscosity of the bulk
electrodes in the model calculations. We did not claim any
quantitative accuracy of our estimates in Ref.~\cite{SZVD}.

In realistic nanoscale structures, the current density may vary
rapidly in the transverse direction due to a decrease of velocity
from the center of the channel to the sides of the conductors. The
transverse density and current density gradients can thus contribute
significantly to the dissipation effects. The contribution can be
further enhanced in the presence of turbulent eddies near the
contacts~\cite{SBHD,DD}. To capture these gradient contributions,
one needs to evaluate the dissipation power $dE/dt = - \int e
\vec{j}\cdot\vec{E}_{xc}d\vec r$ directly (and the associated
resistance as $R^{dyn}=(dE/dt)/I^2$, with $I$ the total current),
because the nonconservative nature of the dynamical xc field makes
it, in general, ambiguous to evaluate a line integral as in
Ref.~\cite{SZVD}. If, once again, the current density and viscosity
are assumed constant, the correction to the resistance evaluated
from the power dissipated is given by $R^{dyn} =
\frac{\eta}{e^2A^2}\int \left[\frac{4}{3}(\partial_z n^{-1})^2 +
(\partial_\perp n^{-1})^2\right]d\vec r$, where $\perp$ represents
the transverse direction, $\eta$ the (constant) viscosity, and $A$
is the cross section of the nanostructure. This expression contains
a positive transverse density gradient term that has been neglected
in Eq. (14) of Ref.~\cite{SZVD}. This term thus increases the
dynamical effects evaluated in Ref.~\cite{SZVD}. More generally,
the transverse density and current density gradients and the spatial
variation of the viscosity must all be taken into account when
evaluating the viscous resistance.  For a general current-carrying
nanoscale system, therefore, a quantitative evaluation of these
corrections requires knowledge of their microscopic current and
density distributions~\cite{SBHD}, and the dissipation power (and
associated resistance) must be evaluated numerically.

2) To calculate the viscosity coefficient, Jung {\it et al.}
have applied the Conti-Vignale formula, Eq. (4.10) of Ref.~\cite{CV}.
This formula is more accurate than the high-density formula we have
used in \cite{SZVD}. However, unlike what Jung {\it el al.} suggest,
the Conti-Vignale formula does {\em not} interpolate between the
high-density and low-density limits of the homogeneous electron
liquid. The simple reason for this is that the {\em exact}
low-density limit of the viscosity of the electron gas is {\em
unknown}. The formula instead comes from a fit to numerical results
of Nifos\'i {\it et al.}, which are based on mode-mode coupling
theory~\cite{NCT}. This theory is certainly not exact in the
low-density limit~\cite{CV}. 
Indeed, due to
the strong correlation effects in the low-density electron gas, where
the electrons are on the verge of
crystallization, it is reasonable to suspect that the relative
viscosity $\eta/n$ might increase well above Eq. (4.10) of ~\cite{CV}
with decreasing density.

We conclude that the comment of Jung {\it et al.}, while
interesting, should not be taken as an indication that the viscosity
corrections to the conductance of real nanoscale structures are 
small. A more accurate treatment of the density and current density
distribution and of the electronic correlations may yield much larger
corrections in realistic systems.

The work was supported by Department of Energy grant DE-FG02-05ER46204. 

\vspace{10pt}
\noindent Na Sai,$^1$ Michael Zwolak,$^2$ Giovanni Vignale,$^3$ and \\
Massimiliano Di Ventra$^4$

\small
\vspace{5pt}
\begin{minipage}{1.5in}
\noindent$^1$Department of Physics, The University of Texas, Austin, Texas 78712, USA

\vspace{5pt}
$^3$Department of Physics \& Astronomy, University of
Missouri, Columbia, Missouri 65211, USA
\end{minipage}
\hspace{0.1in}
\begin{minipage}{1.6in}
$^2$Physics Department, California Institute of Technology, Pasadena, California 91125, USA

\vspace{4pt}
$^4$Department of Physics, University of California, San
Diego, La Jolla, California 92093, USA
\end{minipage}

\end{document}